\documentclass[12pt]{article}

\def\gtwid{\mathrel{\raise.3ex\hbox{$>$\kern-.75em\lower1ex\hbox{$\sim$}}}}
\def\ltwid{\mathrel{\raise.3ex\hbox{$<$\kern-.75em\lower1ex\hbox{$\sim$}}}}
\def\square{\kern1pt\vbox{\hrule height 1.2pt\hbox{\vrule width 1.2pt\hskip 3pt
   \vbox{\vskip 6pt}\hskip 3pt\vrule width 0.6pt}\hrule height 0.6pt}\kern1pt}

\begin{document}

\begin{titlepage}

\begin{flushright}
BRX-TH 589 \\
UFIFT-QG-07-03
\end{flushright}

\vskip 2cm

\begin{center}
{\bf Nonlocal Cosmology}
\end{center}

\vskip 2cm

\begin{center}
S. Deser$^*$
\end{center}

\begin{center}
\it{California Institute of Technology, Pasadena, CA 91125 and \\
Department of Physics, Brandeis University, Waltham, MA 02254}
\end{center}

\begin{center}
and
\end{center}

\begin{center}
R. P. Woodard$^{\dagger}$
\end{center}

\begin{center}
\it{Department of Physics, University of Florida, Gainesville, FL 32611}
\end{center}

\vspace{1cm}

\begin{center}
ABSTRACT
\end{center}
We explore nonlocally modified models of gravity, inspired by quantum loop
corrections, as a mechanism for explaining current cosmic acceleration. These
theories enjoy two major advantages: they allow a delayed response to cosmic
events, here the transition from radiation to matter dominance, and they avoid
the usual level of fine tuning; instead, emulating Dirac's dictum, the required
large numbers come from the large time scales involved. Their solar system
effects are safely negligible, and they may even prove useful to the black
hole information problem.

\begin{flushleft}
\end{flushleft}

\begin{flushleft}
$^*$ e-mail: deser@brandeis.edu \\
$^{\dagger}$ e-mail: woodard@phys.ufl.edu
\end{flushleft}

\end{titlepage}

\section{Introduction}

A variety of complementary data sets \cite{data}
have led to general agreement that the universe is accelerating as if
it had critical density, comprised of about $30\%$ matter and
$70\%$ cosmological constant \cite{analysis}. There is,
however, no current compelling explanation for either the smallness of
$\Lambda$, or for its recent dominance in cosmological history \cite{Lambda}. 
Two existing classes of models, scalars \cite{scalars} and ``$f(R)$'' 
modifications of gravity \cite{NO}, can be arranged to reproduce the observed 
(or any other) expansion history \cite{TW,SRSS,RPW,NO}.
However, neither has an underlying rationale nor do they avoid fine tuning 
\cite{Bludman}. Quantum scalar effects, depending on a very small mass 
have also been proposed \cite{Parker}.

In this work, we account for the current phase of acceleration through
nonlocal additions to general relativity. Such corrections arise naturally
as quantum loop effects and have of course been studied, though in other
contexts \cite{TW,nonloc,SW}. As we will see, even the simple models we 
explore here can both generate large numbers without major fine tuning and 
deliver a delayed response to cosmic transitions, in particular to that 
from radiation to matter dominance at $z \sim 2300$. We will neither 
attempt to derive our models from loop corrections nor to survey generic 
candidates here. Instead, we will show that natural nonlocal operators 
such as the inverse d`Alembertian can explain the time lag between $z \sim 
2300$ and the onset of acceleration at redshift $z \sim 0.7$, without 
recourse to large parameters. Large numbers come in our models precisely
from the long time lags themselves, a mechanism reminiscent of some old
ideas of Dirac.

\section{Nonlocal Triggers}

For simplicity, we deal with homogeneous, isotropic and spatially flat
geometries
\begin{equation}
ds^2 = -dt^2 + a^2(t) d\vec{x} \cdot d\vec{x} \; , \label{FRW}
\end{equation}
These correspond the following Hubble and deceleration parameters
\begin{equation}
H(t) \equiv \frac{\dot{a}}{a} \qquad , \qquad q(t) \equiv
-\frac{a \ddot{a}}{\dot{a}^2} = -1 -\frac{\dot{H}}{H^2} \; , \label{Hubdec}
\end{equation}
and to Ricci scalar\footnote{Our conventions are $R \equiv g^{\mu\nu} 
R_{\mu\nu}$ and $R_{\mu\nu} \equiv \partial_{\rho} \Gamma^{\rho}_{~\nu\mu} 
- \partial_{\nu} \Gamma^{\rho}_{~\rho\mu} + \Gamma^{\rho}_{~\mu \sigma} 
\Gamma^{\sigma}_{~\nu\rho} - \Gamma^{\rho}_{~\nu\sigma} \Gamma^{\sigma}_{~\rho
\mu}$.}
\begin{equation}
R = 6 (1 - q) H^2 \; . \label{Ricci}
\end{equation}
For much of cosmic history $a(t)$ grows as a power of time
\begin{equation}
a(t) \sim t^s \qquad \Longrightarrow \qquad H(t) = \frac{s}{t} \qquad ,
\qquad q(t) = \frac{1-s}{s} \; .
\end{equation}
Perfect radiation dominance corresponds to $s = \frac12$, and perfect
matter dominance to $s = \frac23$. The Ricci scalar of course vanishes
for $s=\frac12$ and is positive for $s = \frac23$. It is the lowest
dimension curvature invariant, and the only simple curvature invariant
to vanish at finite $s$, so we concentrate here on $R$-based models.

We seek the inverse of some differential operator to provide the required
time lag between the transition from radiation dominance to matter dominance
at $t_{eq} \sim 10^5$ years. The simplest choice is the scalar wave operator,
suggested also by the fact that, for our background (\ref{FRW}), dynamical
gravitons obey the scalar wave equation \cite{Grish} with
\begin{equation}
\square \equiv \frac1{\sqrt{-g}} \, \partial_{\rho} \Bigl(\sqrt{-g} \,
g^{\rho\sigma} \partial_{\sigma}\Bigr) \longrightarrow -\frac1{a^3}
\frac{d}{dt} \Bigl( a^3 \frac{d}{dt}\Bigr) \; .
\end{equation}
Acting on any function of time $f(t)$, its retarded inverse reduces to
simple integrations:
\begin{equation}
\Bigl[\frac1{\square} f\Bigr](t) \equiv \mathcal{G}[f](t) = - \int_0^t dt' 
\frac1{a^3(t')} \int_0^{t'} dt'' a^3(t'') f(t'') \; . \label{genf}
\end{equation}

If we make the simplifying (and numerically justified) assumption that the 
power changes from $s=\frac12$ to some other value at $t = t_{eq}$, the 
integrals in (\ref{genf}) are easily carried out for our choice of $f = R$
\begin{equation}
\mathcal{G}[R](t) \Bigl\vert_{s} = -\frac{6 s (2s-1)}{(3s-1)}
\Biggl\{ \ln\Bigl(\frac{t}{t_{eq}}\Bigr) - \frac1{3s -1} + \frac1{3s-1}
\Bigl(\frac{t_{eq}}{t}\Bigr)^{3s-1}\Biggr\} . \label{gfac}
\end{equation}
For the matter dominance value of $s = 2/3$, and at the present time of
$t_0 \sim 10^{10}$ years, this yields
\begin{equation}
\mathcal{G}[R](t_0) \Bigl\vert_{s=\frac23} \simeq -14.0 \; .
\end{equation}
If we think of correcting the field equations by this term (apart from small
additions that enforce conservation, and whose form we will shortly exhibit) 
times the Einstein tensor, this result already illustrates how nonlocality 
allows simple time evolution to generate large numbers without fine tuning.

Much larger values can be obtained through other operators, for example,
the Paneitz operator arising in the context of conformal anomalies \cite{SD}. 
When specialized to our geometry (\ref{FRW}) it takes the form
\begin{equation}
\frac1{\sqrt{-g}} \, \Delta_P \equiv \square^2 + 2 D_{\mu} \Bigl(
R^{\mu\nu} - \frac13 g^{\mu\nu} R\Bigr) D_{\nu} \longrightarrow
\frac1{a^3} \frac{d}{dt} \Bigl(a \frac{d}{dt} a \frac{d}{dt} a 
\frac{d}{dt} \Bigr) \; .
\end{equation}
One gets about $10^6$ from the dimensionless combination of the inverse of
this operator acting on $R^2$.

\section{Specific Models}

Here we evaluate the consequences of the simplest alteration of the
Einstein action,
\begin{equation}
\Delta \mathcal{L} \equiv \frac1{16 \pi G} \, R \sqrt{-g} \times
f\Bigl(\mathcal{G}[R]\Bigr) \; . \label{DL2}
\end{equation}
One could modify the cosmological term in a similar way, but that turns
out to require fine tuning to delay the onset of acceleration sufficiently.

Naively varying a nonlocal action such as (\ref{DL2}) would result in
advanced Green's functions as well as the retarded ones (\ref{genf})
we desire. However, because conservation only depends on the Green's
function being the inverse of a differential operator, one gets causal
and conserved equations by simply replacing the advanced Green's
functions by the retarded ones \cite{SW}.\footnote{To derive causal and
conserved field equations from quantum field theory one uses the
Schwinger-Keldysh formalism \cite{RJ}. This will generally result in
dependence upon the real part of the propagator, as well as the retarded
Green's function, which, if anything, may lead to even stronger effects than
those we consider.} The resulting correction to the Einstein tensor is
\begin{eqnarray}
\lefteqn{\Delta G_{\mu\nu} = \Bigl[ G_{\mu\nu} +
g_{\mu\nu} \square - D_{\mu} D_{\nu} \Bigr] \Biggl\{ f\Bigl(\mathcal{G}[R]
\Bigr) + \mathcal{G}\Bigl[ R f'\Bigl(\mathcal{G}[R]\Bigr)\Bigr]
\Biggl\} } \nonumber \\
& & \hspace{2cm} + \Bigl[\delta_{\mu}^{~(\rho} \delta_{\nu}^{~\sigma)} -
\frac12 g_{\mu\nu} g^{\rho\sigma}\Bigr] \partial_{\rho} \Bigl(\mathcal{G}[R]
\Bigr) \partial_{\sigma} \Biggl(\mathcal{G}\Bigl[ R
f'\Bigl(\mathcal{G}[R] \Bigr) \Bigr] \Biggr) \; . \qquad \label{mod2}
\end{eqnarray}
As promised, it takes the form of a nonlocal distortion of the Einstein
tensor, plus additional terms which enforce the Bianchi identity for any
$g_{\mu\nu}$. The additional terms involve derivatives, so they are
typically small when $f(x)$ varies slowly. Note also that, except for the 
very special case of $f(x)=-x$, no model of this form can be obtained from 
integrating out a scalar. Whatever these models' origin, then, they are not 
scalar-tensor gravities in disguise.

Now note from (\ref{gfac}) that $\mathcal{G}[R](t)$ is small for a long
time after the onset of matter dominance. During this period we may think of
$\Delta G_ {\mu\nu}$ as a perturbation of the stress tensor source, with
$\Delta G_ {00} = -8\pi G \Delta \rho$ and $g^{ij} \Delta G_{ij} = -24 \pi
G \Delta p$. Our corrections will tend to induce acceleration if evolution
during matter domination carries us to the point where
\begin{equation}
\Delta G_{00} + g^{ij} \Delta G_{ij} = -8 \pi G \Bigl(\Delta \rho + 3
\Delta p\Bigr) > 6 q H^2 = \frac43 \cdot \frac1{t^2} \; . \label{Etrace}
\end{equation}
Naturally, once our corrections exceed the Einstein range, they are no longer
perturbations and numerical integration of the field equations is required.

One illustrative class of models has
\begin{equation}
f(x) = C e^{-\frac34 k x} \; .
\end{equation}
The resulting modification $\Delta G_ {\mu\nu}$ gives
\begin{equation}
\Delta G_{00} + g^{ij}
\Delta G_{ij} \simeq \frac43 \cdot \frac1{t^2} \times C \Bigl(1 + \frac34 k
\Bigr) \Bigl(2 - 3 k\Bigr) \Bigl(\frac{t}{t_{eq}}\Bigr)^k .
\end{equation}
Note that the right-hand side is positive for $k$ in the range $-\frac43 <
k < + \frac23$; actually, the range $0 < k < +\frac23$ is needed to make
the correction term grow. Our results do depend on two dimensionless
coupling constants, $C$ and $k$, but neither need be very different from
unity to provide a suitable delay for the onset of acceleration. For example, 
taking $k = .1$ and $C = .2$ would result in about the right onset time, in
accord with the usual meaning of no fine tuning as involving parameters $\sim1$.

As stated earlier, it is possible to construct the scalar potential
$V(\phi)$ to support an arbitrary expansion history $a(t)$ obeying
$\dot{H} > 0$ \cite{TW,SRSS}, and a similar construction exists for 
$f(R)$ theories \cite{NO,RPW}. The same possibility is of course present
in our models, and indeed a procedure has recently been worked out for 
reconstructing the nonlocal distortion function $f(x)$ which would support 
an arbitrary expansion history \cite{DW}. Hence there are certainly 
models of the type (\ref{DL2}) that fit the supernova data. Nor must one
even resort to exotic choices of $f(x)$. As might have been guessed from 
viewing these models as effective nonlocal distortions of Newton's
constant, quiescence at recombination requires that $f(x)$ be small for
$x$ near zero, whereas obtaining de Sitter expansion at asymptotically
late times requires that $f(x)$ approach $-1$ from above for large,
negative $x$. The onset of acceleration is controlled by the range of $x$
at which $f(x)$ becomes of order $-1$.

\section{Conclusions}

We have explored the cosmological effects of some very simple nonlocally
modified Einstein models inspired by loop corrections. Since their actual
derivations from realistic quantum effects are likely to require
nonperturbative summations, we regard them as purely phenomenological for
now. Their two -- equally important -- main virtues are (unlike local
variants): they naturally incorporate a delayed response to the transition 
from radiation to matter dominance, yet avoid major fine tuning. There are
of course many other open questions raised by the present proposal, such as
finding optimal candidate actions while ensuring that nonlocality has no
negative unintended consequences. Some apparent worries, such as (unwanted) 
solar system effects, are easily allayed. There, $\mathcal{G}[R] \sim 
GM/(c^2 r)$ is a small number. Although a single power of $\mathcal{G}[R]$
is observable --- and constrains Brans-Dicke theory tightly \cite{Will} 
--- higher powers, such as occur here, are negligible. 

It should also be mentioned that nonlocality may have a positive use in
the black hole information problem \cite{Steve}: The infalling matter
that creates or accretes to a black hole is imprinted on the external 
geometry through its stress tensor. Nonlocal dependence on the Einstein 
tensor will retain that information; while $T_{\mu\nu}$ does not completely
subsume the matter's internal structure, it is a significant repository 
thereof; furthermore, $\mathcal{G}$ is singular on null surfaces such as 
the event horizon.

\vskip 1cm

\centerline{\bf Acknowledgements}

This work was partially supported by National Science Foundation grants
PHY04-01667, PHY-0244714 and PHY-0653085, and by the Institute for
Fundamental Theory at the University of Florida.

\end{document}